\documentclass[twocolumn,preprintnumbers,
endnote,nofootinbib,prl]{revtex4}
\usepackage{graphicx}

\newcommand{\vev}[1]{\langle {#1} \rangle}
\newcommand{\lsim}{\lesssim}

\newcommand{\eq}[1]{Eq.~(\ref{#1})}

\newcommand{\ord}[1]{\mathcal{O}{(#1)}}
\newcommand{\beq}{\begin{equation}}
\newcommand{\eeq}{\end{equation}}
\newcommand{\bea}{\begin{eqnarray}}
\newcommand{\eea}{\end{eqnarray}}

\newcommand{\mP}{M_{\rm P}}

\begin{document}

\pagestyle{plain}

\title{\boldmath Dark matter repulsion could thwart direct detection}

\author{Hooman Davoudiasl
\footnote{email: hooman@bnl.gov}
}

\affiliation{Department of Physics, Brookhaven National Laboratory,
Upton, NY 11973, USA}


\begin{abstract}

We consider a feeble repulsive interaction between ordinary matter and dark matter, with a range similar to or larger than the size of the Earth.  Dark matter can thus be repelled from the Earth, leading to null results in direct detection experiments,  regardless of the strength of the short-distance interactions of dark matter with atoms.  Generically, such a repulsive force would not allow trapping of dark matter inside astronomical bodies.  In this scenario, accelerator-based experiments may furnish the only robust signals of asymmetric dark matter models, which typically lack indirect signals from self-annihilation.  Some of the variants of our hypothesis 
are also briefly discussed.

\end{abstract} \maketitle

The evidence for cosmic dark matter (DM) has been growing over the last several decades.  Observations based on the gravitational effects of DM have determined that it constitutes about $27\%$ of the energy budget in the Universe \cite{Ade:2015xua}.  However, so far, all attempts to discover DM  through its possible non-gravitational interactions have failed.  In particular, 
direct detection experiments that rely on the recoil of the 
target atomic nuclei or electrons from ambient DM have 
not found any indisputable signals \cite{Tan:2016zwf,Akerib:2016vxi}.  
At the same time, signals of DM annihilation in space have 
not been established \cite{Fermi-LAT:2016uux,TheFermi-LAT:2017vmf}.  Neither, are there any signals of DM annihilation from a trapped population inside 
the Sun \cite{Aartsen:2016zhm} or the Earth \cite{Aartsen:2016fep}.  

One may potentially explain the lack of annihilation signals in 
the case of asymmetric DM \cite{Kaplan:2009ag}, where a conserved charge does not allow such processes (for reviews of this subject see, {\it e.g.}, 
Refs.~\cite{Davoudiasl:2012uw,Petraki:2013wwa,Zurek:2013wia}).  However, some level of interaction between the constituents of ordinary atoms and DM particles is generally assumed for their early Universe production.  Hence, signals in direct detection experiments can be expected to be potentially accessible.

In this work, we propose a scenario that could make it unlikely to detect DM directly in nuclear or electron recoil experiments.  Our basic point is that if there is a long range repulsive interaction 
between ordinary matter - baryons or electrons - and DM, the latter can be repelled from the Earth and other sufficiently large astronomical bodies.  Hence, terrestrial experiments would not be able to detect the recoil signal from ambient Galactic 
DM (Refs.~\cite{Chuzhoy:2008zy,McDermott:2010pa} considered expulsion of DM from the Galactic disk due to a tiny DM electric charge, also leading to null direct detection results.  See also Ref.~\cite{Foot:2014osa} for a different mechanism to shield detectors from the DM wind, based on self-interacting DM.).  In our scenario, it is also unlikely that any 
significant population of DM particles could be captured by stars or planets, eliminating further potential signals of DM.  If DM abundance is set by an asymmetry, it is generally expected that 
there would be no annihilation signals after the early Universe production era.  In this case, accelerator experiments that look for dark sector particles may provide the only feasible access to DM and its associated states (The possibility of new long range forces was introduced in 
Ref.~\cite{Lee:1955vk}.  A review of the subject is presented in Ref.~\cite{Dolgov:1999gk}.  For other preceding work considering possible effects of long range DM interactions, see for example Refs.~\cite{DeRujula:1989fe,Dimopoulos:1989hk,Friedman:1991dj,Gradwohl:1992ue,Gubser:2004uh,Gubser:2004du,Nusser:2004qu,Kesden:2006vz,Kesden:2006zb,Farrar:2006tb,
Kaloper:2009nc,DDDM,Foot:2014uba}).

Let us assume that there is a force mediated by a light boson $X$ that couples to both atoms and DM.  We will assume that this force is repulsive and has a long but finite range.  For the purposes of this work we will assume, unless otherwise specified, that the mediator mass $m_X \sim 10^{-14}$~eV corresponding to a range of $\sim 3 \,R_\oplus \sim 2\times 10^4$~km, where $R_\oplus\approx 6.4 \times 10^3$~km is the radius of the Earth.  One can assume larger ranges, corresponding to smaller $m_X$.  In any event, we will not consider ranges much larger than the size of the Solar system, in order to avoid possible conflict with large scale observations.

A new force with the above range that interacts with ordinary matter is constrained to have a 
tiny coupling $g_{om} \lsim 10^{-24}$ \cite{Schlamminger:2007ht}.   However, the corresponding coupling to DM is not 
constrained at the same level and we choose $g_{dm} \sim 10^{-5}$ as a reference value, which is allowed by all known constraints on DM at scales of interest, as discussed below.  
Let us then estimate  
the potential $\Phi$ generated by all matter in the Earth.  The Earth contains $N_\oplus \sim 10^{51}$ nucleons.  
We will not, at this point, specify how $X$ couples to atoms, but it could be through baryons or electrons.  In either case, the above number gives the correct order of magnitude.  Assuming a repulsive force, we have    
\beq
\Phi(r) \sim g_{om}N_\oplus \frac{e^{-m_X r}}{4 \pi\, r}, 
\label{Phi}
\eeq
where $r$ is the distance to the center of the Earth.  Hence, a DM particle 
would have potential energy 
\beq
V(r)\sim g_{dm} \Phi(r)
\label{V}
\eeq
at a distance $r$ from the center of the 
Earth.  Since we are assuming $m_X\lsim R_\oplus^{-1}$, 
close to the Earth the DM particle will have {\it positive} potential energy 
\beq
V(R_\oplus) \sim  \text{MeV} 
\left(\frac{N_\oplus}{10^{51}}\right)
\left(\frac{g_{om}}{10^{-25}}\right)
\left(\frac{g_{dm}}{10^{-5}}\right).
\label{VR}
\eeq

The virial velocity of DM in our Galactic neighborhood 
is $v_G\sim 10^{-3}$.  Hence, the kinetic energy of DM of mass $m_{dm}$ is given by 
\beq
E_{kin}\sim \frac{1}{2} m_{dm} v_G^2 \sim  500~\text{eV} 
\left(\frac{m_{dm}}{\text{GeV}}\right).
\label{Ekin}
\eeq
For our reference parameters, we find that $E_{kin}\ll V(R_\oplus)$.  
Hence a typical DM particle traveling toward the Earth could be 
repelled back in our scenario, unless it is quite massive.  

We can estimate how massive a DM particle needs to be before it has sufficient kinetic energy to 
reach the Earth in spite of our assumed repulsive force.  To do so, we need to have an upper bound on $g_{dm}$, which we will discuss next.

For a sufficiently long range force, as we have assumed, the cross section 
can be estimated by \cite{Ackerman:mha}
\beq
\sigma_{dm} \sim \frac{g_{dm}^4}{v_{dm}^2 q^2},
\label{sigmadm}
\eeq
where $v_{dm}$ is the typical velocity 
and $q \sim m_{dm} v_{dm}$ is the typical magnitude of DM momentum in the relevant system.

The self interactions of DM  
can be constrained from collisions of galaxy clusters, leading to a bound on 
the corresponding cross section  $\sigma_{dm}/m_{dm} \lsim 1\text{~cm}^2/\text{g}$ \cite{Randall:2007ph}.   Ref.~\cite{Buckley:2009in} considered a conservative bound of
\beq 
\sigma_{dm}/m_{dm} \lsim 0.1\text{~cm}^2/\text{g}\,,
\label{sigbound}
\eeq
from the observations of dwarf galaxy halos.  Given the lower typical velocity of DM in such systems, this can yield a more stringent bound on $g_{dm}$.  Assuming $v_{dm} \sim 10^{-4}$ as a typical value for the dwarf galaxies, we find that the bound in \eq{sigbound} then roughly yields
\beq
g_{dm}\lsim 10^{-3} \left(\frac{m_{dm}}{\text{GeV}}\right)^{3/4}.
\label{gdm-coll-bound}
\eeq

More recently, Ref.~\cite{Agrawal:2016quu} has considered constraints on the a long range force acting on DM.  They find that observations of ``collapsed structures'' - such as dwarf galaxies -  place 
more stringent bounds on $g_{dm}$ than considerations of ``delayed kinetic decoupling'' that could potentially affect large scale structure.  Their results suggest that in the range of DM masses near the 
weak scale $g_{dm}\lsim 10^{-2}$ avoids conflict with astrophysical observations.  
We find that choosing $g_{dm}$ according to the relation in \eq{gdm-coll-bound} is then quite consistent with all existing bounds for values of $m_{dm}$ considered in our discussion.  

Equation (\ref{Ekin}) suggests that even 
for $m_{dm}\sim 400$~TeV - just above the unitarity limit for thermal relic DM
\cite{Griest:1989wd} - we have $E_{kin}\sim 200$~MeV, while \eq{VR} yields $V(R_\oplus) \lsim 10$~GeV for $g_{dm}\sim 0.1$, a value of DM coupling well within the allowed parameter space for such large values of $m_{dm}$ \cite{Agrawal:2016quu}.  Hence, we conclude that for a broad class of particle candidates, our scenario can provide efficient repulsion of DM from the Earth, and hence null results in direct detection experiments.

The above considerations suggest that the repulsion mechanism we have proposed does not 
allow for the accumulation of DM inside the  Earth.  This is also the case for stellar objects, in particular the Sun.  Let us again assume a range of order 
$R_\oplus$, corresponding to a fraction $\ord{10^{-6}}$ of the Solar volume.  
Given that this is roughly the ratio of the masses of the Earth and the Sun, the volume will 
contain at least as much ordinary matter as the whole Earth (taking into account the growing density profile towards the center).  Hence, once the DM particle 
enters the Sun and gets within a distance of $\sim R_\oplus$ of its center, it would be repelled.   
The same conclusion will also hold for white dwarfs and neutron stars, given their solar level masses and much larger densities.  Therefore, we would not expect any major population of DM to gather inside the Sun or other stellar objects.  However, in our scenario, DM can still be present in the Solar system and its neighborhood (unlike the case corresponding to the models in Refs.~\cite{Chuzhoy:2008zy,McDermott:2010pa}).   

We will consider two classes of DM: (i) asymmetric and (ii) thermal relic.  If DM is asymmetric, it generically carries a conserved 
charge and we do not expect 
indirect annihilation signals.  Because of the assumed repulsive force, any other effects from accumulation within astronomical 
bodies are also excluded, as discussed above.  As the repulsion mechanism will lead to 
null direct detection results, the discovery of microscopic 
non-gravitational interactions for asymmetric DM may only 
be possible at accelerator based experiments \cite{FTBD,Alexander:2016aln}.  
Given that asymmetric DM may naturally be  
assumed to be light, with $m_{dm}\sim \text{few}$~GeV \cite{Kaplan:2009ag,Davoudiasl:2012uw,Petraki:2013wwa,Zurek:2013wia}, those experiments will provide well-suited probes in this case.

For asymmetric DM, one may assume that DM and ordinary matter carry charges of the same sign under a gauged long range $U(1)$ whose carrier is identified as the above $X$ boson.  For parameters  as in \eq{VR}, typical of our scenario, assuming that the DM charge $Q_{dm}=1$, one would get $g_{dm}=g_{om}\sim 10^{-5}$, however the charge of ordinary matter must then be chosen to be extremely tiny, $Q_{om}\sim 10^{-20}$.  For a $U(1)$ force, charges are in principle arbitrary, though such an extreme disparity of charges under the same interaction would be puzzling.  

Another possibility is to assume two different $U(1)$ interactions.  For example, one can invoke a gauged $B-L$ number, with $B$ and $L$ baryon and lepton charges, respectively.  Here, we further assume that this interaction is mediated 
by a massless gauge boson, which would be consistent with $g_{om}\sim 10^{-24}$ \cite{Heeck:2014zfa}.  The other $U(1)$ is assumed to be mediated by the $X$ boson of mass $m_X\sim 10^{-14}$~eV, coupled to DM with $g_{dm}\sim 10^{-5}$.  If the two $U(1)$ gauge fields can kinetically mix \cite{Holdom:1985ag}, with a mixing parameter of $\varepsilon \sim 0.1$, we will 
then achieve a scenario in which $X$ is coupled to baryons 
with a strength $\varepsilon  \, g_{om}\sim 10^{-25}$.  
With unit charges for matter fields, the couplings relevant for our scenario will then obtain.  Here, $N_\oplus \sim 10^{51}$ corresponds to the number of neutrons in the Earth, as the contributions of 
protons and electrons cancel out. 

Perturbative loop mediated processes are not effective in generating the necessary $\varepsilon$ in the preceding discussion due to the tiny values of the gauge couplings, unless an enormous number of charged states are considered.  However, one could in principle entertain the possibility that the required kinetic mixing descends from an interaction of the form 
\beq
\frac{S}{\mP} F_{B-L}^{\mu\nu}F_{X\, \mu\nu}
\label{mixing}
\eeq 
induced at the Planck scale $\mP$, perhaps through non-perturbative gravity effects.  Here, $S$ is a scalar whose vacuum expectation value 
$\vev{S} \lsim \mP$ generates the mixing, and $F_{B-L, X}^{\mu\nu}$ denote the field 
strength tensors of the two $U(1)$ gauge interactions.  We will not further speculate on this possibility, but we note that the assumed value of $\varepsilon$ is stable against quantum corrections.
  
If the cosmic DM population is composed of particles and anti-particles, as in the above case (ii)  of thermal relic DM, the $U(1)$ model for $X$ will generally not lead to total repulsion.  That is, either the particle or the anti-particle DM will be attracted to ordinary matter, instead of repelled.  Thus, one may expect a separation of DM particle and anti-particle fluxes over distances of 
$\ord{m_X^{-1}}$, which however still allows for direct detection of one of 
the DM components.   Here, to get total repulsion,  
one could consider $X$ to be a scalar with Yukawa couplings to matter and DM.  We note that the 
scalar $X$ possibility can work for {\it both} asymmetric and thermal relic DM.

Let us assume that $X$ couples to nucleons $N$ and DM $\chi$ through $g_{om} X \bar N N$ and 
$g_{dm} X \bar \chi \chi$, respectively.  If $g_{om} g_{dm} <0$, the Yukawa 
interaction between a nucleon and DM will be repulsive.  Therefore, we will end up with the 
same potential energy as in \eq{V} and our repulsion mechanism will be realized as outlined earlier.  We note that the Yuakwa interaction between two nucleons, or two DM 
states, will be attractive and proportional to $g_{om}^2$ or $g_{dm}^2$, respectively (see 
Ref.~\cite{Adler:2000ib} for a more general discussion). 

In this work, we are mainly motivated by phenomenology and do not address theoretical questions related to natural sizes of various parameters.  
In particular, the stability of the small scalar mass $m_X\sim 10^{-14}$~eV  against quantum corrections will not be discussed in any detail.  However, 
we note that given $g_{dm}\sim 10^{-5}$, a cutoff scale contribution of 
$\sim (16\,\pi)^{-2} g_{dm}^2 \Lambda^2$ may imply $\Lambda\lsim 10^{-8}$~eV.  
The success of our proposal requires coherent scattering on length scales of $\ord{R_\oplus}$, 
which corresponds to momentum transfer $q \lsim R_\oplus^{-1}$, with $R_\oplus^{-1} \sim 10^{-14}$~eV.  Hence, $q \ll \Lambda$ and, strictly speaking, our setup can be viewed as a valid effective theory, cutoff at the scale $\Lambda$. 

\underline{Additional observational implications:} Here, we discuss potential signals of our scenario.  The scenario that we have introduced  requires a new long range force acting on 
ordinary matter. Hence, experiments searching for such forces \cite{Schlamminger:2007ht} 
would in principle be able to uncover some of the underlying physics 
assumed in this work.  One could also consider the effect of the 
hypothetical force on DM distribution and small scale structure.  However, it is not clear that 
for the values of $m_X$ considered here the presence of the long range force could be detectable; this question is beyond the scope of our work.  In any event, variants of the above idea may lead to other 
signals, as we will describe below.   

Let us now consider a larger range for the hypothetical force, comparable to or larger than the radius of the Earth's orbit around the Sun $R_{eo}\approx 1.50\times 10^8$~km, corresponding to $m_X\lsim 10^{-18}$~eV.  We roughly have $R_{eo}/R_\oplus \sim 2\times 10^4$, while $M_\odot/M_\oplus \sim 3\times 10^5$, where the mass of the Sun and the Earth are $M_\odot \sim 2\times 10^{33}$~g and $M_\oplus\sim 6\times 10^{27}$~g, respectively.    Therefore we expect the Sun to dominate the potential  felt by DM near the Earth.  Given that the distance to the Sun has a mild per cent level annual modulation, the energetics of DM particles near the Earth will then have similar modulations.  In particular, if the long range force is repulsive, but only leads to  suppression of DM flux at Earth - {\it not total repulsion} - one could expect a somewhat smaller flux around perihelion $R^p_{eo}\approx 1.47\times 10^8$~km (early January) and a somewhat larger flux around aphelion $R^a_{eo}\approx 1.52\times 10^8$~km (early July) impinging on terrestrial targets.  (The Sun as the source of a long-range potential in neutrino oscillation experiments has been considered, for instance, in Refs.~\cite{LRnu,Davoudiasl:2011sz}; the annual modulation of this effect as a function of the Earth-Sun distance was considered in Ref.~\cite{Davoudiasl:2011sz}.)   

Assuming the long range limit (no Yukawa suppression), the variation of the 
potential will be given by 
\beq
\frac{\delta V}{V} = -\frac{\delta r}{r}
\label{delV}
\eeq
and hence $\delta V/V \sim$1-2$\%$ can be expected.  If the potential has 
a Yukawa form and is exponentially sensitive to $r$, one could in principle have larger variations 
\beq
\frac{\delta V}{V}\approx -m_X \delta r.
\label{YdelV}
\eeq
However, if $m_X \gg 1/r$ the Solar contribution will be severely suppressed compared to that of the Earth.  Hence, one can only expect $\ord{1}$ enhancement of the modulation compared to 
that in \eq{delV}.

The above modulations will be in addition to those expected from the DM ``wind" caused the motion of the Earth around the Sun, as the Solar system moves through the Galactic halo \cite{Drukier:1986tm,Freese:2012xd}.  A more detailed examination is required to ascertain whether and how the flux modulation in our scenario could affect the interpretation of direct detection experiments that look for signal modulation from the DM ``wind" \cite{Bernabei:2010mq}.  

We now briefly consider the possibility that the 
long range force could be attractive.  Here, DM would be accelerated and its kinematic distribution around the Earth would be pushed towards larger velocities.  Depending on the interaction range, this effect could be dominated by the Earth or the Sun.  The acceleration could potentially help with detecting low mass DM by pushing a larger fraction of its population above the recoil energy threshold in direct detection experiments.  Again, if the potential 
is sufficiently long range to be dominated by the Sun, we would expect the effect on 
DM kinematics to have annual modulations, corresponding to Earth-Sun distance variations.  Assuming the attractive force to govern the near-Earth DM kinematics, one expects a somewhat 
harder DM flux around perihelion, and a somewhat softer one around aphelion.

To summarize, a repulsive force of a range comparable to the Earth's size, or a few orders of magnitude larger, that couples to ordinary 
atoms and dark matter can repel the latter away from the Earth.  This could lead to null results 
for direct detection experiments, independently of the strength of the microscopic interactions of 
dark matter particles with atoms.  We noted that the repulsion will exclude accumulation of dark matter inside astronomical bodies, leading to further paucity of signals in this scenario.  In the case of asymmetric dark matter, the lack of indirect annihilation 
signals due to charge conservation will typically limit detection prospects to accelerator based experiments.  Discovery of long range forces acting on 
ordinary matter will lend support to our hypothesis.  Also, if the force has sufficiently long range, the Sun could dominate the potential felt by dark matter near the Earth.  In this case, annual modulations of dark matter kinematics, correlated with the 
Earth-Sun distance, would be expected and could potentially provide another handle on this scenario.

\acknowledgments

We would like to thank Rouven Essig, Eder Izaguirre, Ian Lewis, Bill Marciano, and Chris Murphy for discussions and comments.  This work is supported by the United States Department of Energy under Grant Contract DE-SC0012704.




\clearpage

{\bf Notice}: This manuscript has been co-authored by employees of Brookhaven Science Associates, LLC under Contract No. DE-SC0012704 with the U.S. Department of Energy. The publisher by accepting the manuscript for publication acknowledges that the United States Government retains a non-exclusive, paid-up, irrevocable, world-wide license to publish or reproduce the published form of this manuscript, or allow others to do so, for United States Government purposes.

This preprint is intended for publication in a journal or proceedings.  Since changes may be made before publication, it may not be cited or reproduced without the author’s permission.

{\bf Disclaimer}:
This report was prepared as an account of work sponsored by an agency of the United States Government.  Neither the United States Government nor any agency thereof, nor any of their employees, nor any of their contractors, subcontractors, or their employees, makes any warranty, express or implied, or assumes any legal liability or responsibility for the accuracy, completeness, or any third party’s use or the results of such use of any information, apparatus, product, or process disclosed, or represents that its use would not infringe privately owned rights. Reference herein to any specific commercial product, process, or service by trade name, trademark, manufacturer, or otherwise, does not necessarily constitute or imply its endorsement, recommendation, or favoring by the United States Government or any agency thereof or its contractors or subcontractors.  The views and opinions of authors expressed herein do not necessarily state or reflect those of the United States Government or any agency thereof.


\end{document}